\documentclass[10pt,conference]{IEEEtran}

\IEEEoverridecommandlockouts
\usepackage[utf8]{inputenc}
\usepackage[T1]{fontenc}
\usepackage[ngerman, english]{babel}
\usepackage{csquotes}
\usepackage[style=ieee,sortcites=true,sorting=none,maxbibnames=6, backend=biber,mincitenames=1,maxcitenames=2, dashed=false]{biblatex}
\usepackage{graphicx}
\usepackage{amsmath}
\usepackage{amssymb}
\usepackage{amsthm}
\usepackage{mathtools}
\usepackage{braket}
\usepackage{hyperref}
\usepackage{titlesec}
\usepackage{balance}

\usepackage{amsfonts}
\usepackage{algorithmic}
\usepackage{textcomp}
\usepackage{xcolor}

\def\BibTeX{{\rm B\kern-.05em{\sc i\kern-.025em b}\kern-.08em
    T\kern-.1667em\lower.7ex\hbox{E}\kern-.125emX}}

\graphicspath{{plots}}

\title{Recommending Solution Paths for Solving Optimization Problems with Quantum Computing \thanks{This project is supported by the Federal Ministry for Economic Affairs and Climate Action on the basis of a decision by the German Bundestag through the project QuaST.}}
\author{
    \IEEEauthorblockN{Benedikt Poggel\IEEEauthorrefmark{1}, Nils Quetschlich\IEEEauthorrefmark{2}, Lukas Burgholzer\IEEEauthorrefmark{3}, Robert Wille\IEEEauthorrefmark{2}\IEEEauthorrefmark{4}, Jeanette Miriam Lorenz\IEEEauthorrefmark{1}}
    \IEEEauthorblockA{\IEEEauthorrefmark{1}Fraunhofer Institute for Cognitive Systems IKS, Munich, Germany}
    \IEEEauthorblockA{\IEEEauthorrefmark{2}Chair for Design Automation, Technical University of Munich, Germany}
    \IEEEauthorblockA{\IEEEauthorrefmark{4}Software Competence Center Hagenberg GmbH (SCCH), Hagenberg, Austria}
    \IEEEauthorblockA{\IEEEauthorrefmark{3}Institute for Integrated Circuits, Johannes Kepler University Linz, Austria}
     \{benedikt.poggel, jeanette.miriam.lorenz\}@iks.fraunhofer.de\\ \{nils.quetschlich, robert.wille\}@tum.de, lukas.burgholzer@jku.at
}
\date{January 2023}

\theoremstyle{definition}
\newtheorem*{example}{Example}

\usepackage[capitalize,noabbrev,nameinlink]{cleveref} 

\usepackage{fancyhdr}

\fancypagestyle{specialfooter}{%
  \fancyhf{}
  
  \fancyfoot[R]{ \noindent\fbox{%
    \parbox{\textwidth}{%
        {\footnotesize \copyright 2023 IEEE. Personal use of this material is permitted. Permission from IEEE must be obtained for all other uses, in any current or future media, including reprinting/republishing this material for advertising or promotional purposes, creating new collective works, for resale or redistribution to servers or lists, or reuse of any copyrighted component of this work in other works.}
        }
    }}
}

\begin{document}

\maketitle
\thispagestyle{specialfooter}

\begin{abstract}
Solving real-world optimization problems with quantum computing requires choosing between a large number of options concerning formulation, encoding, algorithm and hardware. Finding good solution paths is challenging for end users and researchers alike. We propose a framework designed to identify and recommend the best-suited solution paths in an automated way. This introduces a novel abstraction layer that is required to make quantum-computing-assisted solution techniques accessible to end users without requiring a deeper knowledge of quantum technologies. State-of-the-art hybrid algorithms, encoding and decomposition techniques can be integrated in a modular manner and evaluated using problem-specific performance metrics. Equally, tools for the graphical analysis of variational quantum algorithms are developed. Classical, fault tolerant quantum and quantum-inspired methods can be included as well to ensure a fair comparison resulting in useful solution paths. We demonstrate and validate our approach on a selected set of options and illustrate its application on the capacitated vehicle routing problem (CVRP). We also identify crucial requirements and the major design challenges for the proposed abstraction layer within a quantum-assisted solution workflow for optimization problems.
\end{abstract}


\section{Introduction}

Optimization problems are ubiquitous in many business fields, from logistics over smart factories to energy network maintenance. Transformed into mathematical models, answering all of these questions is notoriously difficult. Efficient classical algorithms can only find approximations.

Since Grover's algorithm~\cite{grover_fast_1996}, quantum computing (QC) is discussed as an option to improve and outperform classical methods in the field of mathematical optimization. However, the presently available Noisy Intermediate-Scale Quantum (NISQ)~\cite{preskill_quantum_2018} devices are limited in size and affected by noise. Only small quantum circuits can be executed before decoherence destroys the result of the calculation~\cite{gonzalez-garcia_error_2022}. Consequently, algorithms such as Grover's cannot be executed. For near-term quantum advantage, interest has turned towards variational quantum algorithms~\cite{cerezo_variational_2021} whose smaller circuits are suited for NISQ devices. In particular, the variational quantum eigensolver (VQE)~\cite{peruzzo_variational_2014} and the quantum approximate optimization algorithm (QAOA)~\cite{farhi_quantum_2014} have been explored in many variations.

However, realizing the disruptive potential of QC in a \mbox{real-world} setting has proven to be difficult. The pursuit of quantum advantage with NISQ devices needs improvements both on the hardware and software side: Manufacturers find ways to simultaneously scale their qubit systems and reduce noise, and quantum software engineers need to find out what circuits should be executed on the devices to produce a tangible benefit. These questions are currently addressed manually without any broader automated solution.

Optimizing the quantum-assisted solution process is a hard task that goes beyond the algorithm selection. It starts with the selection of appropriate business use cases and ends with guaranteeing that the solution returned by the quantum hardware actually results in an improvement. Every step along the way needs to be fine-tuned to optimally use the available devices. Naturally, considering all available methods requires expertise from various fields. Economists help to identify use cases where an incremental improvement will be most impactful. Mathematicians cast the problem into a well-defined model. Quantum and classical software engineers as well as physicists design and select the best hybrid quantum-classical algorithm. Physicists and engineers attempt to overcome the challenges posed by noise and to improve connectivity. 

Classical computing is not fundamentally different, but a lot can be automated and masked from the end users. With this work, we contribute to creating such an abstraction layer for quantum computing by automating and optimizing the decision steps sketched above. Ultimately, this significantly lowers the barrier of entry and facilitates the application of quantum-enhanced methods. The margin of error is small. Even routines that only add polynomial overhead can quickly negate all advantage. Similarly to quantum machine learning, one should broaden the view from a purely scaling-oriented perspective thinking in complexity classes such as P or NP and strive for improvements in solution quality or generalization capabilities as well~\cite{schuld_machine_2021}.

We envision a modular, semi-automated framework that aims to combine all relevant expertise into well-defined solution paths. Based on different metrics, solution paths are recommended. New methods can be integrated seamlessly. Their performance can be evaluated with application-specific metrics on a per-instance level with a focus on whether one can expect an actual benefit. Even for methods that currently cannot beat the best classical approaches, the framework identifies critical points where an improvement is needed and, thus, allows to estimate a prospective timeline when and under which conditions a reevaluation may be beneficial. Furthermore, the framework can be extended beyond NISQ methods to fault-tolerant and quantum-inspired solution options.

The remaining paper is organized as follows: \Cref{sec:preliminaries} describes the current status of QC-assisted optimization methods and introduces the capacitated vehicle routing problem (CVRP). \Cref{sec:related_work} treats connections to related work and tools. Then, the core concept is presented in \Cref{sec:concept} with a special emphasis on how to compare solution paths in \Cref{sec:metrics}. Our instantiation is detailed in \Cref{sec:realization} including an in-depth treatment of the CVRP example. \Cref{sec:conclusions} summarizes our contribution and concludes with some general remarks and future directions. 

\section{Preliminaries}
\label{sec:preliminaries}
\pagestyle{plain}

\subsection{Solving Optimization Problems with Quantum Computing}

Optimization problems are a well-established application of classical computing and can broadly be stated as ``From a set of alternatives, find the option that optimizes a given objective and at the same time fulfills a set of given constraints". For \mbox{NP-hard} problems, the amount of available options typically grows exponentially with the problem size. This renders provably optimal solution methods infeasible and introduces the need for heuristic methods.

Fault-tolerant quantum computing is expected to yield a quadratic speedup~\cite{grover_fast_1996} for the \emph{unstructured search problem}. With NISQ devices, the hope is that variational algorithms can already realize parts of this advantage. They encode the solution to the problem as the ground state of a quantum Hamiltonian and then aim to find this state via a variational loop~\cite{cerezo_variational_2021}. The standard example is the QAOA~\cite{farhi_quantum_2014} that mimics an adiabatic evolution using a parameterized circuit. Beyond gate quantum computing, quantum annealing is a quantum-inspired protocol specifically tailored to optimization problems~\cite{falco_numerical_1988, johnson_quantum_2011}. 

To date, there is no definite proposal on how to realize this potential for real-world use cases, in part due to the difficulty of optimizing the various choices one is required to make throughout the process. A solution part starting with a business application needs to cast it into a mathematical model, then find an encoding into a quantum system and choose an algorithm to find its ground state. Due to the highly experimental status of the technology, the room for errors in all of these decisions is small and how to realize a practical quantum advantage applications is an open research question.

\begin{figure}[t]
    \centerline{\includegraphics[width=1.05\columnwidth]{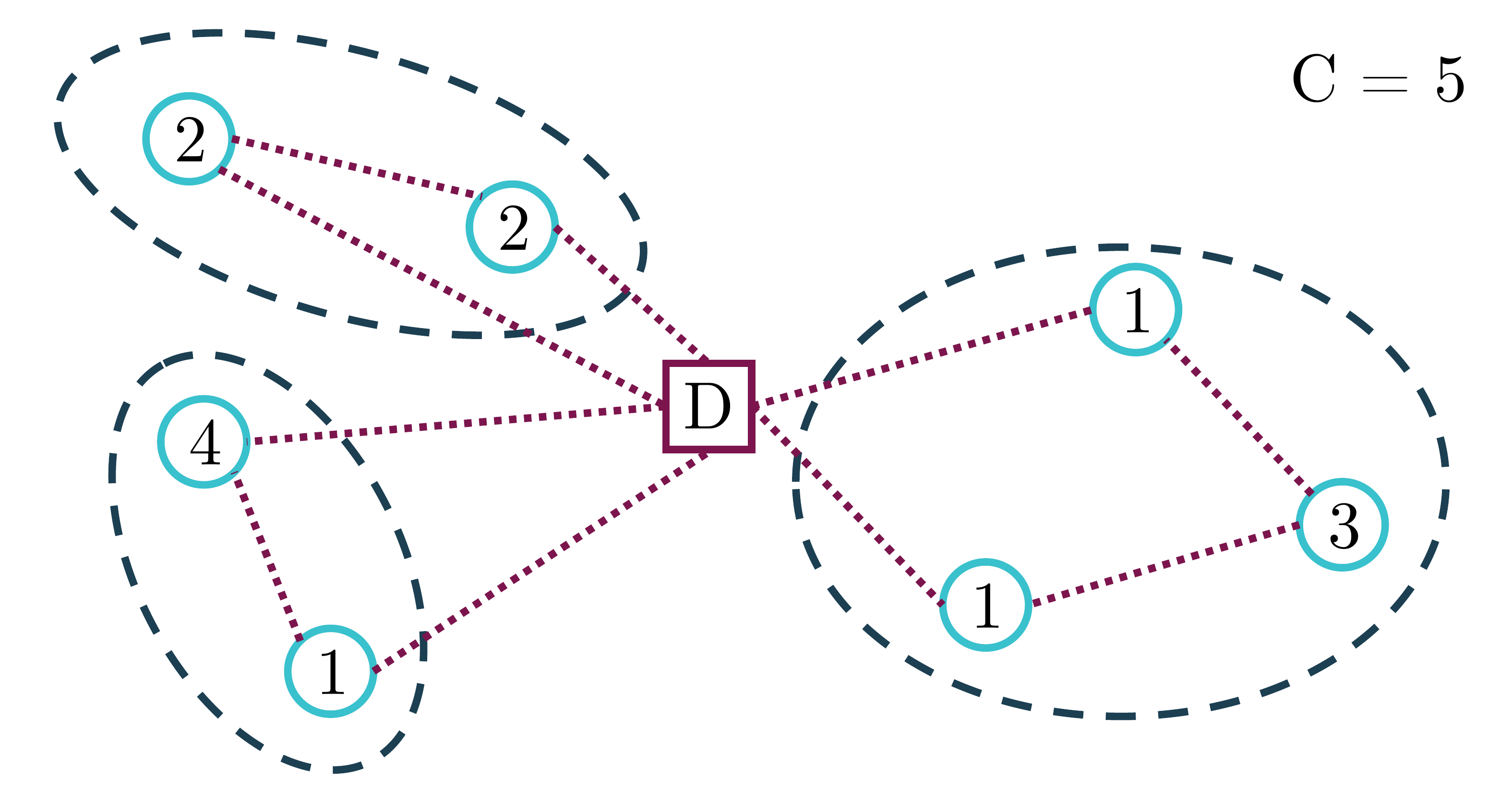}} 
    \caption{An example instance of the CVRP. The depot is marked with the red square. The demand at the circular blue nodes needs to be satisfied with a vehicle of capacity $C=5$. An optimal solution consists of three tours with a shortest path within each tour. \vspace{-0.5cm}}
    \label{fig:cvrp}
\end{figure}

\subsection{An Example Use Case: \\ The Capacitated Vehicle Routing Problem}
\label{sec:cvrp}

The CVRP (see \cref{fig:cvrp}) is an extremely relevant generalization of the travelling salesperson problem (TSP). The task is to collect goods at a number of sites with a vehicle of finite capacity. The vehicle can be emptied again by visiting a special site, the depot. A solution is a collection of multiple routes that all start and end at the depot. Each single route then corresponds to a TSP. However, the full CVRP is considerably more difficult due to the capacity constraint and minimization of the total path length. Formally, the problem is defined on a $n+1$-node complete graph where one node is marked as the depot and all other nodes are equipped with a \emph{demand} $c_i$. The edge weights between the nodes correspond to the distance between the sites. A vehicle with total capacity $C$ should now visit all non-depot nodes once on a set of routes each starting and ending at the depot. The total path length of those routes should be minimal.

Real-world examples are a postal service collecting packages at pickup points or a producer delivering goods to a number of supermarkets. A more thorough analysis of quantum-assisted solution methods for the CVRP is given in~\cite{palackal_quantum-assisted_2023}.

\section{Related Work}
\label{sec:related_work}
The problem of encoding a domain-specific problem in a quantum circuit is an active research area and several works with different foci are already available.
A prominent approach aims to benchmark different QC solutions for the same initial problem to be solved.
In \cite{Finzgar_2022}, a framework is proposed that explores the search space of how to create and execute a problem-specific quantum circuit.
For two sample problems, different paths from the initial problem formulation to an actual QC solution are evaluated. The decisions along those paths are selected and compared based on a brute-force scheme.
Another tool named the quantum solver follows a similar approach~\cite{escanez-exposito_quantumsolver_2022}.
Here, a user can explore the search space from a predefined problem set to different quantum solutions.
This approach is extended further to \mbox{application-specific} approaches such as in \emph{Tangelo}~\cite{tangelo} for chemistry problems.
For all these tools, no informed recommendations are given how to solve a specific problem without executing certain paths to a quantum solution.

On the other hand, instruments have been proposed focusing on automating these problem-to-solution flows without executing and benchmarking all options.
In \cite{quetschlich2022mqtproblemsolver}, conceptual ideas on automating quantum workflows and shielding end users from quantum computing are presented.
Another example is the \emph{MQT Predictor}~\cite{quetschlich2022mqtpredictor} which determines good options how to compile a given quantum circuit on actual QC hardware following a supervised machine learning approach.

\begin{figure*}[ht]
    \centerline{\includegraphics[width=0.9\linewidth]{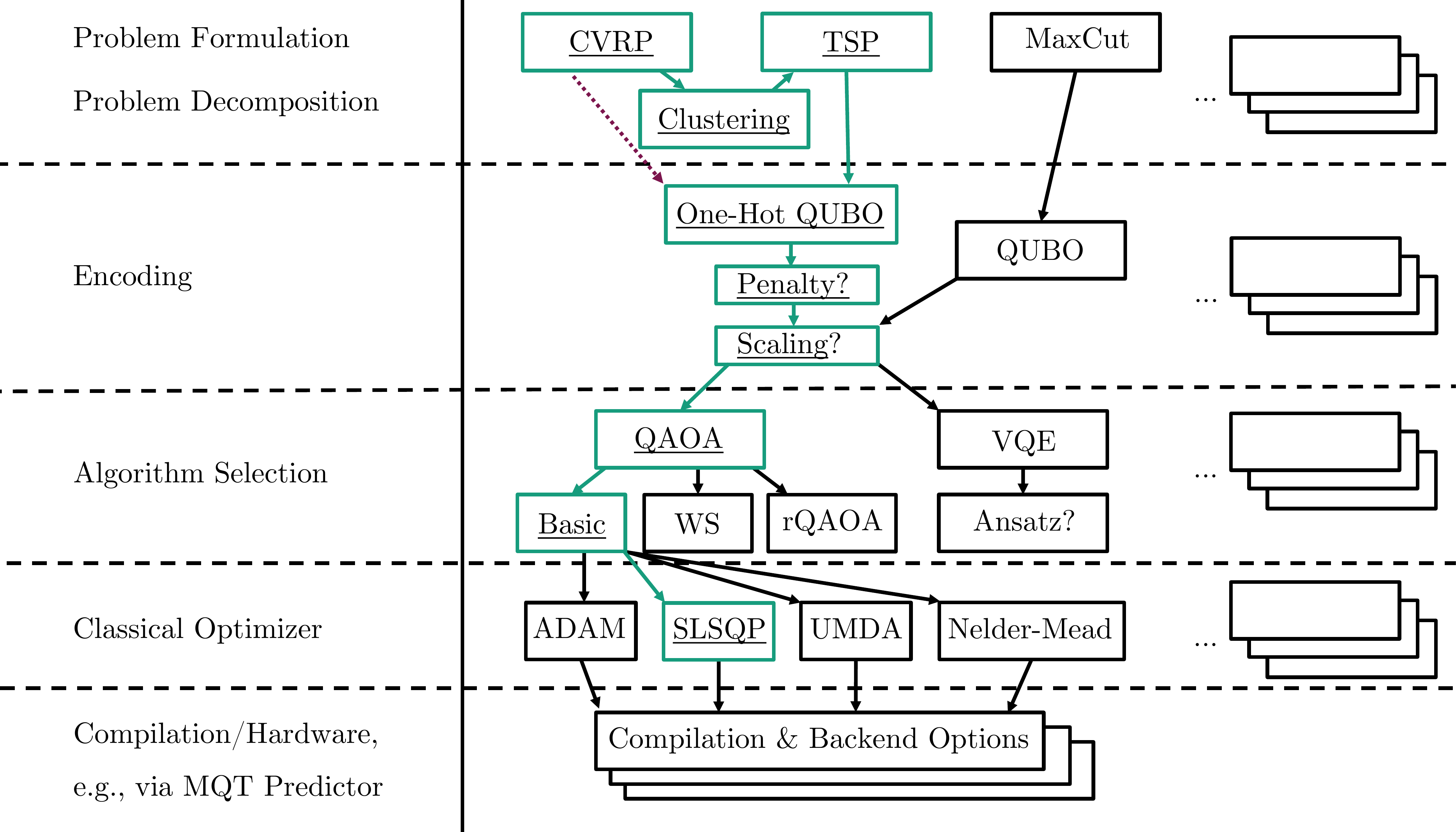}}
    \caption{A sketch of the proposed decision tree with solution paths starting from the top. The left column follows \Cref{sec:concept}. The right column shows the specific options considered in this work with those underlined and marked in green representing a recommended solution path (see \Cref{sec:realization}). The red dotted arrow marks a clearly infeasible path at the moment. Most arrows between the hybrid algorithm and classical optimizer levels have been omitted to avoid confusion.\vspace{-0.5cm}}
    \label{fig: decisiontree}
\end{figure*}

\section{Concept}
\label{sec:concept}

In this work, we start from the actual quantum-independent problem formulation and are concerned with the solution process up to the point where a quantum circuit is created.
There, the circuits can be handed over to a dedicated compilation (recommendation) software.

\subsection{A Modular Decision Tree}

In order to be useful to end users new to QC and to represent the state-of-the-art at the same time, the proposed framework aims to ultimately encapsulate all steps necessary to improve real-world applications with a quantum-assisted solution. Broken down into different levels, the decision points are shown in \cref{fig: decisiontree}:

\begin{itemize}
    \item \emph{Problem Formulation}: The problem identified in a company needs to be formulated as a mathematical optimization problem.
    \item \emph{Problem Decomposition}: Appropriate decomposition techniques should be applied in order to split the problem in parts that can profit from currently available quantum solutions.
    \item \emph{Encoding}: The resulting subproblems need to be encoded into a Hamiltonian or similar, e.g., a quadratic unconstrained binary optimization (QUBO) formulation.
    \item \emph{Algorithm Selection}: In principle, classical, purely quantum and hybrid quantum-classical (variational) algorithms are options to be considered.
    \item \emph{Classical Optimizer}: Since pure quantum solutions are rather distant in the current NISQ era, classical optimizers and algorithms applied in hybrid routines have a crucial influence on the solution quality.
    \item \emph{Compiler and Hardware}: A suitable combination of compiler and hardware needs to be found.
\end{itemize}

It is evident that choosing among the many solution paths spanned by these decision points is no easy task. It requires expertise from mathematics, physics, computer science and even business knowledge. Furthermore, given the current experimental state of QC, there is little room for errors. E.g., a bad optimizer in a variational quantum algorithm quickly fails to find even a bad approximate solution.

To bring QC into the application, an abstraction layer is needed that automatizes the decisions at every level. No single person or group can be expected to cover the entire process optimally though. Therefore, it is crucial that the automated workflow is \emph{modular} and \emph{expandable} such that it can constantly be reviewed and new state-of-the-art proposals can be integrated easily. This requires transparent criteria for the recommended choices as well as precisely defined interfaces that connect the levels. In the end, finding optimal solution paths is a community effort. 

Furthermore, different end users will have various levels of access to quantum hardware and proprietary software tools. Thus, recommendations need to be adaptive to a given user configuration and cannot be hardcoded. This promotes modularity as well and facilitates the inclusion of new methods.

\subsection{How To Find Recommendations}

The criteria used to arrive at recommendations for a specific decision point can be broadly categorized into four types:

\begin{itemize}
    \item \emph{Analytic} insights can provide theoretical performance guarantees, usually under idealized conditions. They include restrictions (e.g., in size) that depend on the quality of the available hardware as well.
    \item \emph{Heuristics} argue by means of intuitive reasoning and analogy, like estimating the effectiveness of a classical optimizer from the structure of the loss landscape.
    \item \emph{Empirical} methods often form the foundation for heuristics, but are a source of knowledge in their own right. In fact, proposals within the scientific community often rely to a great deal on empirical data obtained through numerical simulations.
    \item \emph{Machine learning} (ML) is, in a sense, the automated continuation of empirical findings. It requires foremost a precisely defined value metric that can be used to train a ML network.
\end{itemize}

Clearly, these distinctions are not clear-cut. To find optimal solution paths, one should combine the available knowledge to get the best result. 

Given the complexity of the task and currently missing recommendations or standards, the question remains how to resolve disagreements regarding how to weight arguments concerning the choice of one solution path over another. We plan to tackle this problem in a two-fold way: First, a user with sufficient knowledge should be able to customize all parts of the solution process to their liking and thus deviate from any automatized recommendation. Second, the constant improvement of recommendations will naturally assume a peer review in the form of \emph{propose--review--accept} where the review again is a community effort and the acceptance is based on the acceptance in the expert community.

\section{Monitoring}
\label{sec:metrics}

To address the question of how to compare solution paths and monitor their performance, one needs to define appropriate metrics and tools to support empirical and heuristic arguments. Both should be integrated into the decision finding workflow to efficiently evaluate and integrate new methods and ideas. Furthermore, new evaluation methods and metrics can be discussed and possibly accepted similarly to algorithms in a peer-review manner.

Metrics suited for real-world use cases need to be specific to the problem classes since, e.g., QUBO or Ising approximation ratios do not translate directly into a solution quality such as the TSP path length~\cite{palackal_quantum-assisted_2023}. Assuming the algorithm produces feasible states, the path length can even be used directly. 

 \begin{figure}[ht]
    \centerline{\includegraphics[width=0.95\columnwidth]{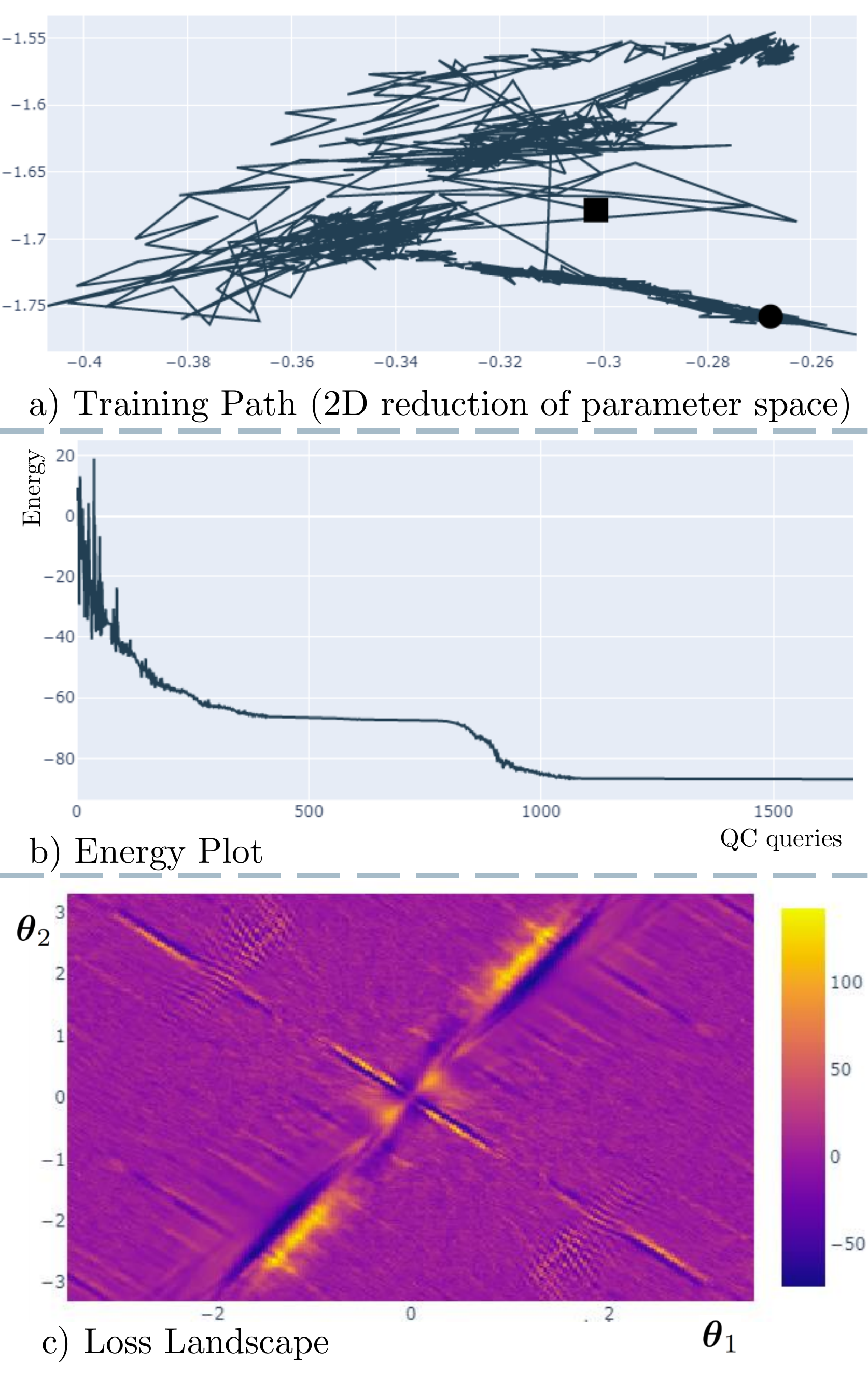}} 
    \caption{The plotting tools included to visualize the performance of a variational algorithm. a) A 2-dimensional representation (here via t-SNE~\cite{tSNE}) of the trajectory in the parameter space. The square and circle mark the initial and final points respectively. b) The energy throughout the optimization procedure. c) The loss landscape over an arbitrary plane in parameter space.\vspace{-0.5cm}}
    \label{fig:plotting_tools}
\end{figure}

In terms of evaluation tools, variational quantum algorithms need to be examined in both their quantum and classical aspects. The classical optimization process can be visualized by plotting the training trajectory in the parameter space (or a 2-dimensional reduction of it) as well as by tracking the Ising energy throughout the process (\Cref{fig:plotting_tools} a) and b)). The quantum ansatz on the other hand defines the loss landscape the classical optimizer operates on (\Cref{fig:plotting_tools}c)). Arguments can also be supported by circuit-centric metrics such as entangling capacity or expressibility~\cite{sim_expressibility_2019}.
\section{Implementation}
\label{sec:realization}

The envisioned workflow has been implemented on a selection of example problems by populating all levels with a subset of options among which an initial recommendation is given. Only hardware and compiler options are not included and can be taken care of with a tool like the MQT Predictor~\cite{quetschlich2022mqtpredictor}. The resulting instance includes all options shown in \cref{fig: decisiontree} in a modular software setup. We demonstrate the workflow throughout all levels from the problem formulation to the circuit. The concrete decision points are described in detail in the following sections and are illustrated by the CVRP for clarity. Despite this division into layers, the difficulty lies in the strong interdependence between these levels. Choosing between the options even at the highest level requires some thought about the target hardware, decomposition methods etc.


\subsection{Problem Formulation \& Decomposition}

For each business use case, a number of different mathematical formulations as well as equivalent formulations among them will be possible. They typically differ in the type of and connection between variables and constraints (assuming the overall optimization goal, e.g., reducing path lengths driven by a vehicle fleet, is fixed and defines the overall use case). In addition, classical preprocessing steps that decompose the problem on the highest level may be applied.

Relevant criteria are -- even at this level -- the number of qubits and their connectivity on the target hardware as well as shifting complexity between the objective function and constraints. Their relationship is nontrivial: For example, even binary variables do not necessarily map onto qubits in a 1-to-1 relation and the realization of their interactions may require a larger or smaller overhead when mapped to the target hardware system.

\begin{example}
For the CVRP, a solution needs to contain two units of information the index of the path it is contained in, and where it is located in said path. A direct encoding is possible but cumbersome~\cite{irie_quantum_2019} and uses decision variables $x_{ijk}$ where the first index corresponds to the route it is contained in, the second to the node and the third to the (discretized) time the node is serviced. As a rough estimate, this formulation requires $\mathcal{O}(n^2)$ qubits where $n$ is the total number of nodes. Since this requirement is out of the reach of current hardware, one has to adopt the 2-phase solution for the CVRP which first performs clustering and then solves a TSP on each cluster~\cite{laporte_heuristics_2002}. The clustering is performed classically with a \mbox{bottom-up} hierarchical approach~\cite{nielsen_hierarchical_2016}. This results in a set of TSP problems for which the direct binary encoding is $\mathcal{O}(n'^2)$~\cite{lucas_ising_2014} with a cluster size $n' < n$.
\end{example}

\subsection{Encoding}

The encoding translates the problem into a form suitable for quantum computing, typically a Hamiltonian and possibly a set of constraints. The current standard path is a QUBO formulation. There are other options, however, including higher-order product terms (HOBO)~\cite{glos_space-efficient_2020} and alternative ways to handle constraints like unbalanced penalization~\cite{alej2022unbalanced}.

For the mapping of integer variables onto qubits, strategies like one-hot, binary and domain-wall encoding~\cite{chancellor_domain_2019} are available, each with their unique advantages and disadvantages. For example, one-hot encoding usually leads to a moderate connectivity and a simple interpretation of individual variables. However, all feasible states lie in local minima since they are separated by at least two bit flips.

When it comes to the cost function, the overall scale of the Hamiltonian can influence the performance of QAOA-like algorithms~\cite{brandhofer_benchmarking_2022}. It should suit the concrete algorithm and the hardware to ensure that the individual gates can be executed as reliably as possible. Furthermore, both problem-inherent and encoding-specific constraint terms need to be penalized. Mathematics dictates a minimum penalty to ensure that the lowest eigenstate corresponds to the solution of the problem, but gives no upper bound. In order to achieve a well-behaved spectrum and optimization landscape, the penalty should not be too large, however.

\begin{example}
For the TSPs resulting from the CVRP clustering, the standard binary encoding with $(n-1)^2$ variables~\cite{lucas_ising_2014} is used which corresponds to the one-hot encoding of the integer TSP formulation. Our focus, then, lies on the automation of the scaling and penalty factor selection. For toy problems, the minimum penalty factor $P_{min}$ can be found by a classical search. It seems sensible to choose the penalty as $P = \lambda P_{min}$ with $\lambda \approx 1.2$. Empirically, $P_{min}$ does not scale with the number of nodes. If $P_{min}$ is not accessible, $P$ can be selected according to a characteristic length of the system, e.g., the side length of the bounding box containing all nodes.

Second, the overall scaling factor, while mathematically irrelevant, can lighten the load on the classical optimizer by shifting the relevant parameters and parameter changes into a reasonable dimension. Since the variational circuit usually consists of basic gates such as Pauli rotations, a sensible approach is to align the spectrum of the $n$-qubit Hamiltonian to that of a Pauli operator. For example,~\cite{brandhofer_benchmarking_2022} proposes to choose the scaling factor such that the spectral width of a QAOA mixer ($2n$ for a standard X mixer) lines up with the spectral width of the Hamiltonian. For toy problems, the Hamiltonian is diagonalizable with the usual numerical methods and this method can be applied directly. For \mbox{real-world} use cases, an estimate for the spectral width is required. Two options are possible. First, Gershgorin circles can be used to obtain an upper bound to the spectral width~\cite{gershgorin_uber_1931}. This upper bound, then, should be scaled with a factor that is yet to be determined to get a realistic estimate. Second, good scaling factors for instances of the same or a similar problem class (e.g., uniformly distributed random nodes within the same bounding box) can be used.
\end{example}

\subsection{Algorithm Selection}

The choice of the hybrid algorithm and its hyperparameters is at the center of the decision tree. Proven approximation ratios, e.g., for MaxCut with QAOA~\cite{farhi_quantum_2014} do not translate directly into the solution quality for many constrained problems such as the TSP. For this reason, the algorithms need to be benchmarked respecting a problem-specific performance metric. Efficient testing involves simulators as a first step, but then also needs to connect to the specific hardware.

Furthermore, algorithm-specific hyperparameters need to be selected. VQE for example needs an ansatz, QAOA a specific number of layers. As a first recommendation, we choose the QAOA depth such that all qubits can be entangled by the problem Hamiltonian evolution according to a reverse causal cone argument~\cite{streif_training_2019, farhi_quantum_2014}.

\begin{example}
For the TSP, qubits are interacting if they are associated the same or an adjacent time step, or the same node. As a baseline, we use the standard QAOA algorithm. Warm-start QAOA has the problem that the variable assignment of any reasonable warm-start solution will typically be very different from the best solution due to the separation of feasible solutions. Recursive QAOA cannot provide any scaling advantage since it still requires to encode the entire problem and suffers from the weakly correlated QAOA output.

Concerning the number of layers, a 2-layer QAOA is in principle enough to couple all interacting qubits in the 4-node problem. Nonetheless, to make sure the resulting circuit does not run into expressibility issues, a 5-layer circuit is employed instead making sure that all qubits can be entangled.
\end{example}

\subsection{Classical Optimizer}

The classical optimizer is both important and universal enough to merit its own level in the decision process. It has a strong impact on the optimization process and influences the likelihood of getting stuck in local minima or on barren plateaus~\cite{mcclean_barren_2018}. In addition, it requires its own hyperparameters like the step size of a gradient routine.

Relevant criteria are the number of circuit executions needed for one iteration as well as the typical number of iterations needed to converge. Furthermore, noise resilience is clearly important for all optimizers, but can still take a larger or smaller role depending on the structure of the loss landscape and noise characteristics of the target device.

\begin{example}
For the considered TSP, four selected optimizers are implemented through Qiskit: ADAM~\cite{kingma_adam_2014}, SLSQP~\cite{kraft_software_1988}, UMDA~\cite{soloviev_quantum_2022} and Nelder-Mead~\cite{nelder_simplex_1965}. These are chosen to broadly cover various optimization strategies, namely gradient-based, gradient-free and genetic algorithms.

By examining simulated training trajectories showing the energy during the optimization process, it becomes clear that UMDA does not achieve good energy values whereas the other three find similar energies at the end of the process. However, ADAM needs more executions of the quantum circuit by an order of magnitude. Nelder-Mead shows a stable, but slow improvement of the energy, but often fails to converge within a reasonable number of maximum iterations. In the end, ADAM will time out on real backends which is why SLSQP is the optimizer of choice for the problem at hand.
\end{example}
\section{Conclusions}
\label{sec:conclusions}

We present the concept of a recommendation framework for the quantum-assisted solution of optimization problems. It acts as an abstraction layer for users that are unfamiliar with a part or all of the necessary decisions. Problem formulation and decomposition, encoding, algorithm and hyperparameter selection, and the compiler and hardware choice need to be carefully interlinked. We describe the various foundations for good recommendations as well as the basic requirements of the framework: a modular, expandable layout, clearly defined interfaces that do not stand in the way of innovations and a setup that can be tailored towards specific end users. We demonstrate our approach for selected example options. For the CVRP use case, we illustrate the application of the proposed framework in detail.

An integrated framework is a cornerstone in order to realize a practical quantum advantage as early as possible. Our application-driven performance metrics stay as close as possible to the use case, but this is not without drawbacks: Modifications of the solution path influence the result not as directly as common circuit-centric measures. Thus, it is challenging to draw concrete conclusions at the specific decision points, especially given the problem-specific character of the performance metrics.

In the scientific community, there is no shortage of new ideas. However, a lot of them target only specific points of the optimization workflow, with end-to-end solutions still wanting. A significant effort is required to ensure transparency and comparability -- crucial conditions for the success of QC.

Finally, working solution paths for applications in the near future require a high degree of tuning at every one of the highly interdependent levels (\Cref{fig: decisiontree}). This makes it particularly hard to design a highly modular decision tree. Creators of new algorithms or techniques that improve steps of the optimization process should not be hindered, but helped. Instead, they need to have a clear path on where to insert their innovation, what requirements it needs to provide, and what line of argumentation is needed to convince the community of its relevance. To this end, we plan to expand and improve the framework into several directions:

First, the options need to be expanded to ultimately cover all relevant approaches. They need to be integrated into the decision tree, tested and compared against the existing methods. The application-driven performance measures will lead to recommendations tailored not only to the problem class, but even to the instance. The tremendous complexity of this endeavor is justified by the improvement in solution quality that is needed to render quantum-assisted methods useful.

Second, many of the current recommendations rely on heuristics. We need to improve our understanding of the underlying assumptions and explore the general connections between characteristics such as circuit depth and \mbox{application-centric} performance measures. Once an impactful and \mbox{well-reasoned} metric is found, techniques based on machine learning become available.

Various tools have been proposed to automatize parts of the solution process (see \Cref{sec:related_work}), but our approach differs in that it includes the entire process and specifically aims to optimize how different levels of the software stack need to work together to produce meaningful results. This enormous undertaking can only succeed as a community effort. For this reason, we are happy to answer questions from interested researchers and research groups and start collaborating.

\balance
\clearpage
\printbibliography

\end{document}